\documentclass[aps,prl,amsmath,twocolumn,superscriptaddress,letterpaper,floatfix]{revtex4-1}
\usepackage{graphicx,color}
\usepackage{verbatim}
\usepackage{amssymb} 
\usepackage{amsmath}
\usepackage{amsfonts}
\usepackage{mathdots}
\usepackage{hyperref}
\usepackage{epsfig}
\usepackage{braket}
\usepackage{bm}

\usepackage{soul}

\begin{document}
	\title{A Unified Theory of Deterministic Magnetic Switching}

    \author{Xizhi Fu}
    \affiliation{Department of Physics, Hong Kong University of Science and Technology, Clear Water Bay, Hong Kong SAR, China}
    
    \author{Lei Han}
    \affiliation{Key Laboratory of Advanced Materials (MOE), School of Materials Science and Engineering, Tsinghua University, Beijing 100084, China}
    
    \author{Xi Liu}
    \affiliation{Department of Physics, Hong Kong University of Science and Technology, Clear Water Bay, Hong Kong SAR, China}
    
    \author{Cheng Song}
    \affiliation{Key Laboratory of Advanced Materials (MOE), School of Materials Science and Engineering, Tsinghua University, Beijing 100084, China}
    
    \author{Junwei Liu}
    \email{liuj@ust.hk}
    \affiliation{Department of Physics, Hong Kong University of Science and Technology, Clear Water Bay, Hong Kong SAR, China}
    \affiliation{IAS Center for Quantum Matter, Hong Kong University of Science and Technology, Clear Water Bay, Hong Kong SAR, China}

    
	\date{\today}
	\begin{abstract}  
    The deterministic switching of magnetic order parameters is critically important, as it forms the fundamental basis for manipulating information states in magnetic memory devices. This work presents a general theoretical framework that unifies the mechanisms of magnetic switching by introducing the concept of \emph{switching symmetry}
    and establishing that the necessary condition for deterministic switching is the breaking of \emph{all} switching symmetries, which can be achieved through asymmetric states, asymmetric barriers, and asymmetric torques. 
    Our theory can successfully and universally explain \emph{all} reported experimental cases of deterministic magnetic switching and provides unified and simple design principles for new switching devices of all magnetic materials without the need of complicated simulations.
	\end{abstract}
	\maketitle
	
\emph{Introduction}---  
The study of the deterministic switching of order parameters in magnetic systems has a long and rich history \cite{hirohata2020}. In ferromagnets (FMs), net magnetization can be reversed by a magnetic field or current-induced torques, such as spin-transfer torques (STTs) \cite{slonczewski1996,berger1996,ralph2008,brataas2012} and spin-orbit torques (SOTs) \cite{miron2011,luqiao2012,luqiao2012_2,fukami2016,zelezny2017,manchon2019,liang2020,liang2022,cai2017,xiao2019,chen2021,wu2025,guo2025}. These fundamental discoveries have propelled the development of magnetoresistive random-access memory, a cornerstone in non-volatile memory technology. 
In contrast, antiferromagnets (AFMs) are expected to exhibit superior performance for next-generation spintronic devices because of their ultrafast dynamics and high integration density \cite{jungwirth2016,baltz2018}. In particular, the recently discovered altermagnets (A$l$Ms) \cite{makoto2019,hayami2019,libor2020,samanta2020,linding2020,haiyang2021,rafael2021,mazin2021,dingfu2021,libor2022,bai2024,yiyuan2024} further integrate the properties and advantages of FMs and conventional AFMs by exhibiting spin-splitting band structures and zero net magnetization in the nonrelativistic limit. 

However, the compensated magnetic moments in conventional AFMs and A$l$Ms make manipulation of the N{\'e}el vector exceptionally challenging. It is only in recent years that electrical switching of the N{\'e}el vector has been demonstrated, and so far only in a limited number of AFMs and A$l$Ms \cite{wadley2016,godinho2018,lei2024,zhou2025,zhou2026,wenqing2024,bodnar2018,chen2018,moriyama2018,duttagupta2020,yan2020,cheng2020,meer2021}; see also related theoretical studies \cite{zelezny2014,xue2021,zhengde2023,shuisen2026}.
Despite these substantial experimental advances, one central question remains unresolved: What are the fundamental requirements for \emph{deterministic switching}, a process in which controllable external stimuli can unambiguously drive the system from any initial state to a uniquely designated final state?

In this work, we introduce the new concept of switching symmetry, which is the symmetry operation that maps distinct magnetic ground states to each other, and demonstrate that the necessary condition for deterministic switching is the breaking of all switching symmetries. This general principle governs switching dynamics across various mechanisms, including asymmetric state energies, asymmetric energy barriers, or asymmetric driving torques.
We rigorously validate this theory by analyzing the role of switching symmetry and external stimuli on switching dynamics based on the Landau–Lifshitz–Gilbert (LLG) equation and further explicitly demonstrate its power in typical FMs, AFMs, and A$l$Ms.
	
\begin{figure}[tb]
\includegraphics[width=1.0\linewidth]{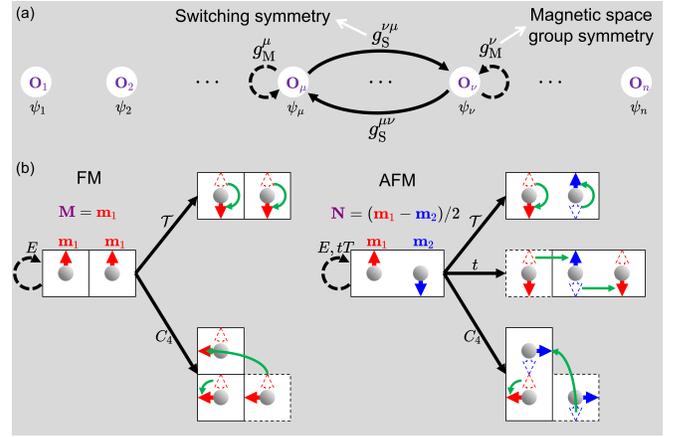}
\caption{\label{fig_1} Schematic illustration of symmetry operations and their effects on ground states.  
(a) In a magnetic system with n-fold degenerate ground states, the switching symmetry $g_\mathrm{S}^{\nu\mu}$ maps the ground state $\psi_\mu$ to $\psi_\nu$, thereby transforming the order parameter from $\mathbf{O}_\mu$ to $\mathbf{O}_\nu$. In contrast, the magnetic space group symmetry $g_\mathrm{M}^{\mu}$ leaves $\psi_\mu$ and $\mathbf{O}_\mu$ invariant.  
(b) 
Left: In a FM with a single sublattice, the identity $E$ preserves the ground state, whereas the time-reversal $\mathcal{T}$ reverses the net magnetization $\mathbf{M} = \mathbf{m}_1$. Right: In an AFM with two sublattices, the N{\'e}el vector $\mathbf{N} = (\mathbf{m}_1 - \mathbf{m}_2)/2$ is reversed under either $\mathcal{T}$ or the translation $t$. The four-fold rotation $C_4$ rotates both $\mathbf{M}$ (FM) and $\mathbf{N}$ (AFM) by 90$^\circ$.}
\end{figure}
			
\emph{Unified theory}--- 
We begin by establishing the unified theory for magnetic switching. A typical magnetic system can be described by the local magnetic moments $\mathbf{m}_i$ at the sites $\mathbf{r}_i$ in a periodic crystal characterized by the crystal space group $\mathrm{G}_\mathrm{C}=\{g_\mathrm{C}\}$. Represented by a classical pseudo-vector function $\mathbf{m}(\mathbf{r}_i)$, each $\mathbf{m}_i$ transforms as $\mathbf{\widetilde{m}}(\mathbf{r}_i)=g_\mathrm{C}\mathbf{m}(g_\mathrm{C}^{-1}\mathbf{r}_i)$ under the $g_\mathrm{C}$ operation \cite{xiao2024,yang2024}, with $\mathbf{m}$ being invariant under inversion ($\mathcal{P}$) and translation ($t$) operations. Without external stimuli, the Hamiltonian $\mathcal{H}(\mathbf{m})$ of magnetic moments generally remains invariant under all $g_\mathrm{C}$ operations and time-reversal ($\mathcal{T}$) operation, i.e., $\mathcal{H}(\mathbf{m})=\mathcal{H}(\mathbf{\widetilde{m}})$, where $\mathbf{m}\equiv[\mathbf{m}(\mathbf{r}_1),\mathbf{m}(\mathbf{r}_2),\ldots,\mathbf{m}(\mathbf{r}_N)]$ and $\mathbf{\widetilde{m}}\equiv[\mathbf{\widetilde{m}}(\mathbf{r}_1),\mathbf{\widetilde{m}}(\mathbf{r}_2),\ldots,\mathbf{\widetilde{m}}(\mathbf{r}_N)]$, and $\mathbf{\widetilde{m}}(\mathbf{r}_i)=g\mathbf{m}(g^{-1}\mathbf{r}_i)$ for $g=g_\mathrm{C}$ or $\mathcal{T}$. Consequently, the complete symmetry group of $\mathcal{H}$ is $\mathrm{G}_\mathcal{H} = \mathrm{G}_\mathrm{C}+\mathrm{G}_\mathrm{C}\mathcal{T} = \{g_\mathrm{C},g_\mathrm{C}\mathcal{T}\}$.

Typically, a magnetic system can undergo a temperature-driven phase transition from the paramagnetic phase to a long-range ordered magnetic ground state $\psi_\mu$, characterized by an order parameter $\mathbf{O}_\mu$ that emerges from the spontaneous breaking of symmetries $g_\mathrm{S}^{\nu\mu} \in \mathrm{G}_\mathcal{H}$. Each $g_\mathrm{S}^{\nu\mu}$ operation maps the ground state $\psi_\mu$ to another ground state $\psi_\nu$ [Fig.~\ref{fig_1}(a)], simultaneously switching the order parameter from $\mathbf{O}_\mu$ to $\mathbf{O}_\nu$, and thus $g_\mathrm{S}^{\nu\mu}$ is termed the \emph{switching symmetry}. The ground state $\psi_\mu$ remains invariant under each operation of unbroken symmetries $g_\mathrm{M}^\mu$, which form the magnetic space group $\mathrm{G}_\mathrm{M}^\mu = \{g_\mathrm{M}^\mu\}$, a subgroup of $\mathrm{G}_\mathcal{H}$ with the cosets $\widetilde{\mathrm{G}}_\mathrm{S}^{\nu\mu} = \{g_\mathrm{S}^{\nu\mu}\}$ \cite{sm}. As exemplified in Fig.~\ref{fig_1}(b), the time-reversal symmetry $\mathcal{T}$ is a general switching symmetry that can facilitate the transition between two ground states with opposite order parameters, such as net magnetization $\mathbf{M}$ in FMs and N{\'e}el vector $\mathbf{N}$ in collinear AFMs. In AFMs, the translation $t$ may also be a switching symmetry that induces the same transition by interchanging magnetic sublattices. More generally, systems with higher degeneracy can possess additional switching symmetries and hence support other switching types, where an example is that a rotation $C_4$ can mediate a 90$^\circ$ reorientation of the order parameter in systems with four-fold degeneracy. 

\begin{figure}[t]
\includegraphics[width=1.0\linewidth]{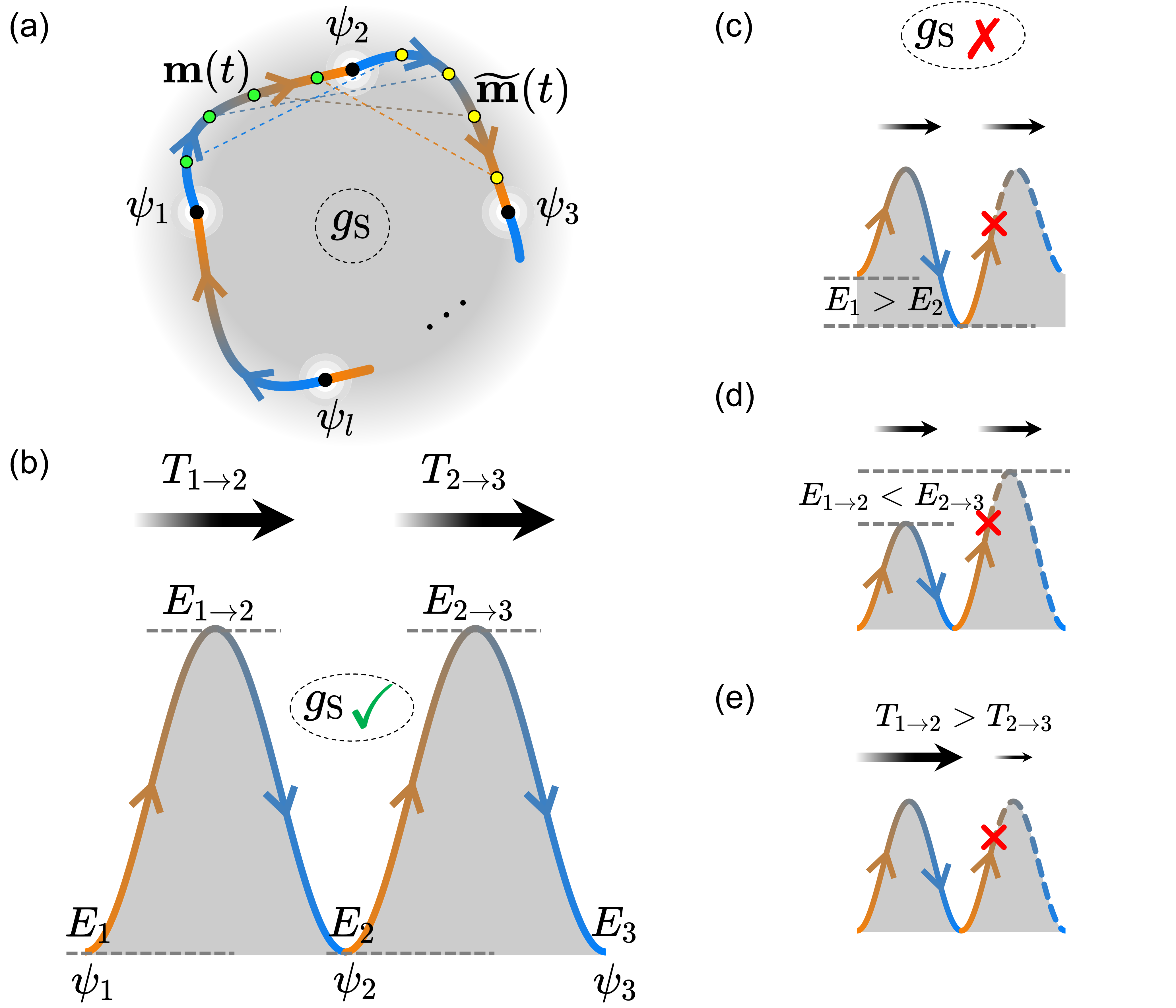}
\caption{\label{fig_2} (a) Phase diagram of transition between ground states $\psi_1,\psi_2,...,\psi_l$, with switching symmetry $g_\mathrm{S}$ connecting intermediate states $\mathbf{m}(t)$ (green points) and $\mathbf{\widetilde{m}}(t)$ (yellow points). (b)-(e) Transition in magnetic systems with:  
(b) unbroken $g_\mathrm{S}$, (c) asymmetric states, (d) asymmetric barriers, and (e) asymmetric torques. In (b)-(e), the curve enclosing the gray region represents the energy profile along the path; solid curves with arrowheads denote allowed paths, while dashed curves with red crosses mark forbidden paths;
black arrows indicate driving torques.}
\end{figure}

In general, switching symmetry can fundamentally prevent deterministic switching by connecting the desired switching path to other symmetry-related paths. To illustrate this point, we consider a switching symmetry $g_\mathrm{S}$ of order $l$, which cyclically permutes the ground states $\psi_1,\psi_2,...,\psi_l$ according to $g_\mathrm{S}\psi_\mu=\psi_{\mu+1}$ ($1\leq\mu\leq l$, $\psi_{l+1} \equiv \psi_1$). As schematically shown in Fig.~\ref{fig_2}(a), besides mapping $\psi_1$ to $\psi_2$, $g_\mathrm{S}$ also maps each intermediate state $\mathbf{m}(t)$ on the switching path $\psi_1\rightarrow\psi_2$ to another intermediate state $\mathbf{\widetilde{m}}(t)$ on the path $\psi_2\rightarrow\psi_3$. More specifically, as illustrated in Fig.~\ref{fig_2}(b), $g_\mathrm{S}$ forces the state energies ($E_1 = E_2$), energy barriers ($E_{1 \rightarrow 2} = E_{2 \rightarrow 3}$), and driving torques ($T_{1 \rightarrow 2} = T_{2 \rightarrow 3}$) to be identical. Suppose external stimuli that preserve $g_\mathrm{S}$ could drive the system to switch from $\psi_1$ to $\psi_2$ along the path $\psi_1\rightarrow\psi_2$, it will equivalently switch $\psi_2$ to $\psi_3$, resulting in a continuous cyclic evolution $\psi_1 \to \psi_2 \to ... \to \psi_l \to \psi_1$. Therefore, we arrive at the central conclusion that a \emph{necessary condition} for deterministic switching in magnetic systems is the breaking of all switching symmetries that connect desired states with undesired states. In particular, this can occur via at least one of three mechanisms: asymmetric states ($E_1 \neq E_2$) [Fig.~\ref{fig_2}(c)], asymmetric barriers ($E_{1 \rightarrow 2} \neq E_{2 \rightarrow 3}$) [Fig.~\ref{fig_2}(d)], and asymmetric torques ($T_{1 \rightarrow 2} \neq T_{2 \rightarrow 3}$) [Fig.~\ref{fig_2}(e)].

To substantiate the above conclusion more rigorously, we now formulate the switching process explicitly in terms of spin dynamics. For a switching process from $\psi_1=\mathbf{m}(t_1)$ to $\psi_2=\mathbf{m}(t_2)$, the final state can be written as
\begin{equation}
\label{eq_path12}
\psi_2=\psi_1+\int_{t_1}^{t_2}\dot{\mathbf{m}}\,\mathrm{d}t .
\end{equation}
The time derivative $\dot{\mathbf{m}}=\frac{\mathrm{d}\mathbf{m}}{\mathrm{d}t}$ is governed by the LLG equation
\begin{equation}  
\label{eq_LLG}  
\frac{\mathrm{d} \mathbf{m}_i}{\mathrm{d} t} = -\gamma \mathbf{m}_i \times \mathbf{H}_i (\mathbf{m}) + \lambda \mathbf{m}_i \times \frac{\mathrm{d} \mathbf{m}_i}{\mathrm{d} t} + \mathbf{T}_i(\mathbf{m}), 
\end{equation}  
where $\gamma$ is the gyromagnetic ratio, $\mathbf{H}_i = -\frac{1}{\mu_0} \frac{\partial \mathcal{H}}{\partial \mathbf{m}_i}$ is the effective field, $\lambda$ is the Gilbert damping parameter, and $\mathbf{T}_i$ is the spin torque exerted by external stimuli. We then ask what happens if the same stimuli are applied for another equal time interval, namely from $t_2$ to $t_3$ with $t_3-t_2=t_2-t_1$.
If the torque preserves the switching symmetry, i.e., $\mathbf{T}_i(\mathbf{m}(t_2+\Delta t))=g_\mathrm{S}\mathbf{T}_j(\mathbf{m}(t_1+\Delta t))$ for all $\Delta t\in[0,t_2-t_1]$, then the LLG equation is covariant under $g_\mathrm{S}$ (see Supplementary Materials \cite{sm} for a derivation). Therefore, since $g_\mathrm{S}\psi_1=\psi_2$, by the uniqueness of the solution with a given initial condition, one obtains $\dot{\mathbf{m}}(t_2+\Delta t)=g_\mathrm{S}\dot{\mathbf{m}}(t_1+\Delta t)$, which implies
\begin{equation}
g_{\mathrm S}\int_{t_1}^{t_2}\dot{\mathbf{m}}\,\mathrm{d}t=\int_{t_2}^{t_3}\dot{\mathbf{m}}\,\mathrm{d}t \nonumber .
\end{equation}
Applying $g_\mathrm{S}$ to Eq.~(\ref{eq_path12}) therefore gives
\begin{equation}
g_\mathrm{S}\psi_2
=g_\mathrm{S}\left(\psi_1+\int_{t_1}^{t_2}\dot{\mathbf{m}}\,\mathrm{d}t\right)
=\psi_2+\int_{t_2}^{t_3}\dot{\mathbf{m}}\,\mathrm{d}t \nonumber .
\end{equation}
Since $g_\mathrm{S}\psi_2=\psi_3$, one has
\begin{equation}
\label{eq_path23_2}
\psi_3=\psi_2+\int_{t_2}^{t_3}\dot{\mathbf{m}}\,\mathrm{d}t .
\end{equation}
This result shows that, once $g_\mathrm{S}$ remains unbroken, the path $\psi_1\rightarrow\psi_2$ necessarily implies the coexistence of the subsequent path $\psi_2\rightarrow\psi_3$. Therefore, deterministic switching from $\psi_1$ to $\psi_2$ requires the applied torque to break every switching symmetry $g_\mathrm{S}$, namely
\begin{equation}
\label{eq_T_gT}
\mathbf{T}_i(g_\mathrm{S}\mathbf{m})\neq g_\mathrm{S}\mathbf{T}_j(\mathbf{m}) .
\end{equation}
In the switching process discussed above, $\mathbf{m}$ and $g_\mathrm{S}\mathbf{m}$ correspond to $\mathbf{m}(t_1+\Delta t)$ and $\mathbf{m}(t_2+\Delta t)$, respectively.

 
In particular, spin torques are commonly decomposed into field-like torques (FLTs, $\bm{\tau}_i^\mathrm{FL}=-\gamma\mathbf{m}_i\times\mathbf{p}_i^{\mathrm{FL}}$) and damping-like torques (DLTs, $\bm{\tau}_i^\mathrm{DL}=-\gamma\mathbf{m}_i\times(\mathbf{m}_i\times\mathbf{p}_i^{\mathrm{DL}})$) \cite{manchon2019,kurebayashi2014}, breaking the switching symmetries $g_\mathrm{S}$ when
\begin{equation}
\label{eq_condition_FLTDLT}
g_\mathrm{S} \mathbf{p}_j^\mathrm{FL} \neq  \mathbf{p}_i^\mathrm{FL} \text{ or } g_\mathrm{S}\mathbf{p}_j^\mathrm{DL} \neq  \mathbf{p}_i^\mathrm{DL}, 
\end{equation}
where $\mathbf{p}_i^\mathrm{FL}$ and $\mathbf{p}_i^\mathrm{DL}$ are polarizations. In general, FLTs can be incorporated into the energy, thus giving rise to asymmetric states and barriers, whereas DLTs are non-conservative and directly contribute to asymmetric torques. Beyond FLTs, polarizations $\mathbf{p}_i$ can modify the Hamiltonian as $\mathcal{H}'=\mathcal{H}+\Delta\mathcal{H}$, with the linear coupling term
\begin{equation}
\label{eq_Hamiltonian_pi}
\Delta\mathcal{H}= \sum_{i} \mathbf{p}^\mathrm{T}_i \mathbf{h}_i \mathbf{m}_i,
\end{equation}
where each $\mathbf{h}_i$ is a 3$\times$3 coupling matrix constrained by $\mathrm{G}_\mathrm{C}$. As a concrete example, for a two-sublattice system ($i=1,2$) with uniform polarizations ($\mathbf{p}_1=\mathbf{p}_2=\mathbf{p}$), a minimal low-order symmetry-allowed form of the Hamiltonian can be written as (see Supplementary Materials \cite{sm} for a derivation)
\begin{eqnarray}
\label{eq_Hamiltonian_total}
\mathcal{H}'&=&\mathcal{H} + h_0 \mathbf{p} \cdot \mathbf{M} + h_{d,N} \mathbf{d} \cdot (\mathbf{p} \times \mathbf{N}) \nonumber\\
& &+ h_{d,M} (\mathbf{p} \cdot \mathbf{d}) (\mathbf{M} \cdot \mathbf{d}) + h_{k,M} (\mathbf{p} \cdot \mathbf{k}) (\mathbf{M} \cdot \mathbf{k}),
\end{eqnarray}
where $\mathbf{d}$ is the Dzyaloshinskii–Moriya (DM) vector, and $\mathbf{k}$ is the easy axis.

\begin{figure*}[htb]
\includegraphics[width=0.85\linewidth]{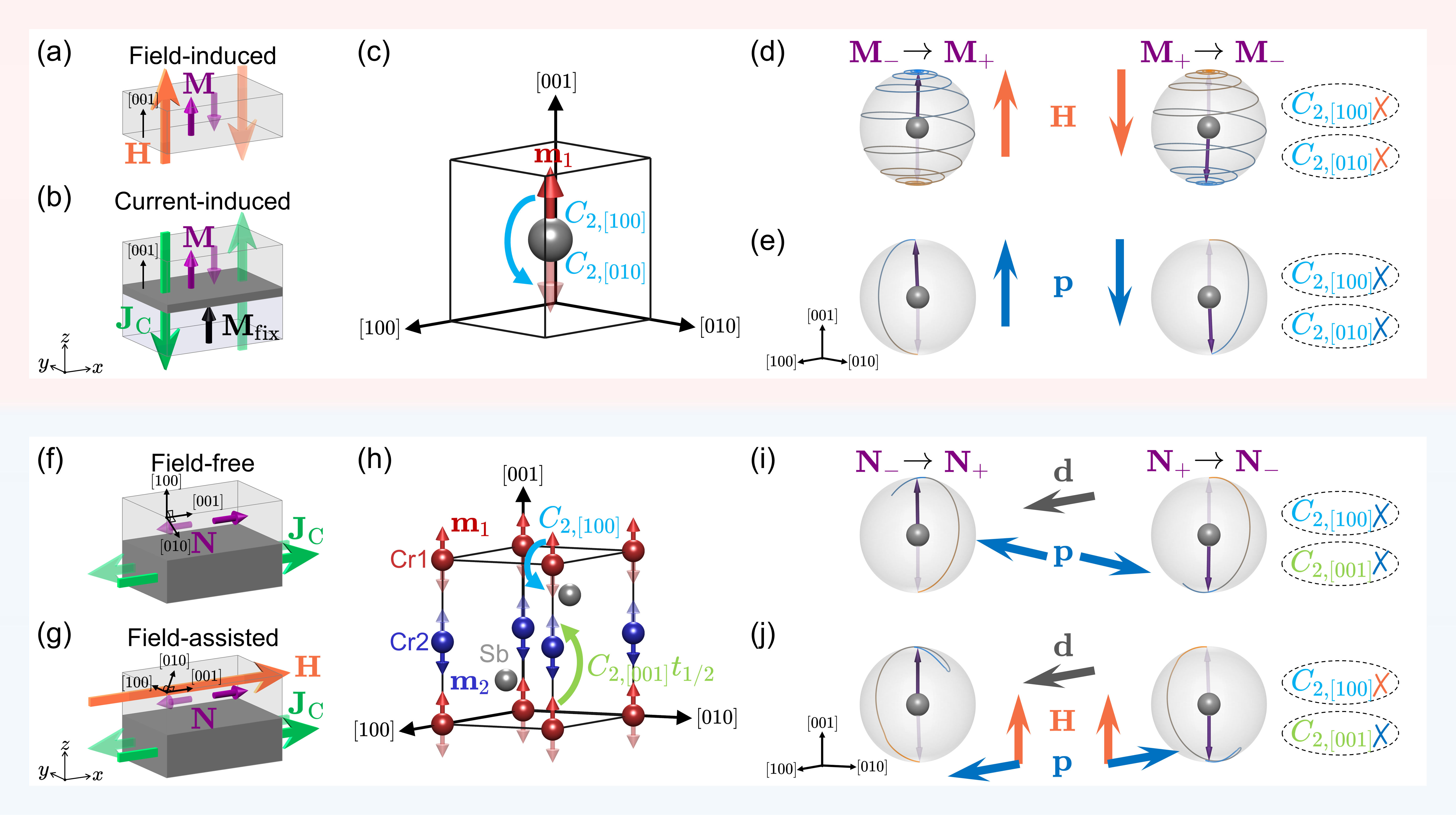}
\caption{\label{fig_3} Switching cases in FMs and A$l$Ms. (a),(b) Schematic diagrams of net magnetization ($\mathbf{M} \parallel\pm\hat{\mathbf{z}}$) reversal in a FM driven by (a) a magnetic field $\mathbf{H}\parallel\pm\hat{\mathbf{z}}$ or (b) a charge current $\mathbf{J}_\mathrm{C}\parallel\mp\hat{\mathbf{z}}$ passing through a fixed FM layer with magnetization $\mathbf{M}_\mathrm{fix}\parallel+\hat{\mathbf{z}}$, where $\mathbf{H}$ and $\mathbf{M}_\mathrm{fix}$ are slightly tilted to initiate the dynamical process. (c) Switching symmetries ($C_{2,[100]}$ and $C_{2,[010]}$) for a single magnetic moment $\mathbf{m}_1$ in an orthorhombic cell. (d),(e) Atomistic simulations of $\mathbf{M}$ reversal between $\mathbf{M}_\pm\parallel\pm[001]$ driven by (d) $\mathbf{H}\parallel\pm[001]$ and (e) a current-induced polarization $\mathbf{p}\parallel\pm[001]$, each with a slight tilt. (f),(g) Schematic diagrams of (f) field-free and (g) field-assisted switching of N{\'e}el vector ($\mathbf{N} \parallel\pm\hat{\mathbf{x}}$) in A$l$M CrSb films (top layers), achieved by $\mathbf{J}_\mathrm{C}\parallel\pm\hat{\mathbf{x}}$ passing through the bottom heavy metal layers. (h) Switching symmetries ($C_{2,[100]}$ and $C_{2,[001]} t_{1/2}$, where $t_{1/2}$ is a half-translation along [001]) for a CrSb thin film with $\mathbf{m}_1$ and $\mathbf{m}_2$. (i),(j) Atomistic simulations of $\mathbf{N}$ reversal between $\mathbf{N}_\pm\parallel\pm[001]$ with DM vector $\mathbf{d}\parallel[100]$ in (i) field-free ($\mathbf{p}\parallel\mp[120]$) and (j) field-assisted ($\mathbf{p}\parallel\pm[100]$ and $\mathbf{H}\parallel[001]$) modes.}
\end{figure*}

\emph{Case studies}--- 
To validate our unified theory, we examine representative switching cases reported in the literature. A canonical example is the net magnetization reversal in a FM driven by a magnetic field [Fig.~\ref{fig_3}(a)] or by a current-induced STT [Fig.~\ref{fig_3}(b)] \cite{ralph2008}. In the macrospin model \cite{xiao2005}, where the FM is described as a single magnetic moment $\mathbf{m}_1$ residing in an orthorhombic cell, the two ground states ($\mathbf{m}_1\parallel\pm[001]$) are connected by the switching symmetries $C_{2,[100]}$ and $C_{2,[010]}$, corresponding to two-fold rotations about $[100]$ and $[010]$ axes, respectively [Fig.~\ref{fig_3}(c)]. According to Eq.~(\ref{eq_condition_FLTDLT}), either a magnetic field $\mathbf{H}\parallel\pm[001]$ or a DLT (from STT) with polarization $\mathbf{p}\parallel\pm[001]$ can break $C_{2,[100]}$ and $C_{2,[010]}$, enabling deterministic reversal of net magnetization, as demonstrated by atomistic simulations [Figs.~\ref{fig_3}(d) and (e)].

Importantly, our theory also provides a unified interpretation of deterministic 180$^\circ$ N{\'e}el vector switching in AFMs and A$l$Ms. As exemplified by A$l$M CrSb thin films \cite{zhou2025}, such switching can be achieved by a charge current that injects an in-plane polarized spin current via the spin Hall effect, enabling both field-free and field-assisted modes for distinct crystallographic orientations [Figs.~\ref{fig_3}(f) and (g)]. As illustrated in Fig.~\ref{fig_3}(h), CrSb possesses two switching symmetries: $C_{2,[100]}$ flipping all magnetic moments and $C_{2,[001]}$ (accompanied by a half-translation) interchanging two sublattices. In the field-free mode, the current-induced spin polarization $\mathbf{p}\parallel\mp[120]$ breaks all switching symmetries, as dictated by Eq.~(\ref{eq_condition_FLTDLT}). Specifically, the direct coupling between $\mathbf{p}$ and $\mathbf{N}$ generates asymmetric states described by the term $h_{d,N} \mathbf{d} \cdot (\mathbf{p} \times \mathbf{N})$ ($\mathbf{d}\parallel[100]$) in Eq.~(\ref{eq_Hamiltonian_total}), thus enabling deterministic $\mathbf{N}$ reversal, as simulated in Fig.~\ref{fig_3}(i). In the field-assisted mode, by contrast, the spin polarization $\mathbf{p}\parallel\pm[100]$ preserves $C_{2,[100]}$. Hence, an auxiliary magnetic field ($\mathbf{H}\parallel[001]$) is required to break this symmetry [Eq.~(\ref{eq_condition_FLTDLT})] and create asymmetric energy barriers, thereby enabling deterministic switching driven by $\mathbf{p}$ [Fig.~\ref{fig_3}(j)].

\begin{figure}[b]
\includegraphics[width=1.0\linewidth]{fig_4.png}
\caption{\label{fig_4} Current-induced 90$^\circ$ and 120$^\circ$ switching of the N{\'e}el vector. (a) A charge current $\mathbf{J}_\mathrm{C}$ ($\mathbf{J}_\mathrm{C}\parallel\pm[100]$ or $\mathbf{J}_\mathrm{C}\parallel\pm[010]$) rotates the N{\'e}el vector $\mathbf{N}$ to be perpendicular to $\mathbf{J}_\mathrm{C}$. (b) A charge current $\mathbf{J}_\mathrm{C}$ ($\mathbf{J}_\mathrm{C}\parallel\pm[210]$, $\mathbf{J}_\mathrm{C}\parallel\pm[120]$, or $\mathbf{J}_\mathrm{C}\parallel\pm[1\bar10]$) rotates the N{\'e}el vector $\mathbf{N}$ to be parallel or antiparallel to $\mathbf{J}_\mathrm{C}$.}
\end{figure}

Apart from deterministic 180$^\circ$ switching of order parameters, our unified theory can also account for other types of switching. For example, a four-fold AFM such as tetragonal CuMnAs \cite{wadley2016} possesses four ground states, characterized by $\mathbf{N}$ aligned along $[100]$, $[\bar100]$, $[010]$, and $[0\bar10]$, respectively. Since anisotropic magnetoresistance cannot distinguish reversed N{\'e}el vectors, we focus on switching between two ground-state classes: $\psi_1$ ($\mathbf{N}\parallel\pm[100]$) and $\psi_2$ ($\mathbf{N}\parallel\pm[010]$). The corresponding switching symmetry set is
\begin{equation}
\label{eq_Gs_90}
\widetilde{\mathrm{G}}_\mathrm{S}^{\mathrm{90^\circ}}=\{C_{4,[001]},C_{2,[110]},C_{2,[1\bar10]},C_{4,[001]}\mathcal{P},M_{[110]},M_{[1\bar10]}\},\nonumber
\end{equation}
where $M$ denotes the mirror symmetry with the normal vector parallel to the subscript. As illustrated in Fig.~\ref{fig_4} (a), a charge current $\mathbf{J}_{\mathrm{C}}$ parallel to $\pm[100]$ or $\pm[010]$ can break all these switching symmetries [Eq.~(\ref{eq_condition_FLTDLT})], enabling deterministic 90$^\circ$ switching of $\mathbf{N}$ between $\psi_1$ and $\psi_2$. In this case, $\mathbf{N}\perp\mathbf{J}_\mathrm{C}$ due to asymmetric states created by current-induced FLTs with staggered spin polarizations \cite{sm}. Similarly, a six-fold AFM (e.g., $\alpha$-$\mathrm{Fe_2O_3}$ \cite{cheng2020}) exhibits three classes of ground states: $\psi_1$ ($\mathbf{N}\parallel\pm[210]$), $\psi_2$ ($\mathbf{N}\parallel\pm[120]$), and $\psi_3$ ($\mathbf{N}\parallel\pm[1\bar10]$), connected by switching symmetry sets
\begin{eqnarray}
\widetilde{\mathrm{G}}_\mathrm{S}^{21},\widetilde{\mathrm{G}}_\mathrm{S}^{12}&=&\{C_{3,[001]},C_{6,[001]},C_{2,[110]},C_{2,[1\bar10]}\},\nonumber\\
\widetilde{\mathrm{G}}_\mathrm{S}^{32},\widetilde{\mathrm{G}}_\mathrm{S}^{23}&=&\{C_{3,[001]},C_{6,[001]},C_{2,[010]},C_{2,[210]}\},\nonumber\\
\widetilde{\mathrm{G}}_\mathrm{S}^{31},\widetilde{\mathrm{G}}_\mathrm{S}^{13}&=&\{C_{3,[001]},C_{6,[001]},C_{2,[100]},C_{2,[120]}\},\nonumber
\end{eqnarray}
where the improper switching symmetries ($C_3\mathcal{P}$, $C_6\mathcal{P}$, and $M$) are omitted for brevity. As illustrated in Fig.~\ref{fig_4} (b), a charge current $\mathbf{J}_\mathrm{C}\parallel\pm[210]$ can break the switching symmetries within $\widetilde{\mathrm{G}}_\mathrm{S}^{12}$ and $\widetilde{\mathrm{G}}_\mathrm{S}^{13}$ [Eq.~(\ref{eq_condition_FLTDLT})], allowing deterministic switching to $\psi_1$ with $\mathbf{N}$ either parallel or antiparallel to $\mathbf{J}_\mathrm{C}$. However, $\mathbf{J}_\mathrm{C}\parallel\pm[210]$ preserves $C_{2,[210]}$ in $\widetilde{\mathrm{G}}_\mathrm{S}^{23}$, so $\psi_2$ and $\psi_3$ remain indistinguishable under such a charge current. Consequently, a charge current oriented along $\pm[210]$, $\pm[120]$, or $\pm[1\bar10]$ can deterministically switch $\mathbf{N}$ by $60^\circ$ or $120^\circ$ to align parallel or antiparallel to $\mathbf{J}_\mathrm{C}$. Microscopically, this behavior can be understood from asymmetric torques offered by current-induced DLTs with uniform spin polarizations \cite{sm}. In addition, more switching cases are discussed in the Supplementary Materials \cite{sm}.

\emph{Discussion and outlook}---
We establish a simple procedure to achieve deterministic magnetic switching: (i) identify all the switching symmetries that connect the target ground states with others; and (ii) design the external stimuli to break all such switching symmetries. Although thermal fluctuations, domain wall motions, lattice distortions, and metastable states may affect the quantitative switching path and threshold, they do not alter the validity of our central symmetry-based principle. This work paves the way for further exploration and innovation in spintronics, particularly for the design of controllable magnetic memory materials.

Our unified theory also offers an insightful perspective on A$l$M. As established in Ref.~\cite{haiyang2021}, A$l$M is defined by the crystal symmetry $C$ instead of $\mathcal{P}$ or $t$ that connects opposite-spin sublattices, which fundamentally allows the nonrelativistic spin currents available for electrical readout of the N{\'e}el vector. Interestingly, this very same symmetry $C$ is precisely the switching symmetry that must be broken to achieve deterministic N{\'e}el vector reversal. Our theory thus reveals a duality in A$l$M spintronics: the symmetry $C$ that shapes the readout is the one whose breaking enables robust writing.

\emph{Acknowledgments}---
This work is supported by National Key R$\&$D Program of China (2021YFA1401500) and the Hong Kong Research Grants Council (CRS$\_$HKUST603/25, C6046-24G, 16306722, 16304523 and 16311125).
	 
	\bibliographystyle{apsrev4-1}
	\bibliography{References}
\end{document}